\documentclass[iop]{emulateapj}
\usepackage{graphicx}
\usepackage{setspace}
\usepackage{natbib}
\usepackage{color}
\usepackage{amsmath,amssymb}
\usepackage{times}
\usepackage{aas_macros}

\begin{document}

\title{The velocity anisotropy of distant Milky Way halo stars from
  \textit{Hubble Space Telescope} proper motions}
\author{A.J. Deason\altaffilmark{1,3}, R.P. Van der
  Marel\altaffilmark{2}, P. Guhathakurta\altaffilmark{1}, S.T. Sohn\altaffilmark{2},T.M. Brown\altaffilmark{2}}

\altaffiltext{1}{Department of Astronomy and Astrophysics, University
  of California Santa Cruz, Santa Cruz, CA 95064, USA; alis@ucolick.org}
\altaffiltext{2}{Hubble Fellow}
\altaffiltext{3}{Space Telescope Science Institute, 3700 San Martin
  Drive, Baltimore, MD 21218, USA}
\date{\today}

\begin{abstract}
Based on long baseline (5--7 years) multi-epoch \textit{HST}/ACS photometry,
used previously to measure the proper motion of M31, we present the proper
motions (PMs) of $13$ main-sequence Milky Way halo stars. The sample lies at an average distance of $r \simeq 24$ kpc from the Galactic center, with a root-mean-square spread of 6 kpc. At this distance, the median PM accuracy is 5 km s$^{-1}$. We devise a maximum likelihood routine to determine the tangential velocity ellipsoid of the stellar halo. The velocity second moments in the directions of the Galactic $(l,b)$ system are $\langle v^2_l \rangle^{1/2} = 123^{+29}_{-23}$ km s$^{-1}$, and $\langle v^2_b \rangle^{1/2} = 83^{+24}_{-16}$ km s$^{-1}$. We combine these results with the known line-of-sight second moment, $\langle v^2_{\rm los} \rangle^{1/2} = 105 \pm 5$ km s$^{-1}$, at this $\langle r \rangle$ to study the velocity anisotropy of the halo. We find approximate isotropy between the radial and tangential velocity distributions, with anisotropy parameter $\beta = 0.0^{+0.2}_{-0.4}$. Our results suggest that the stellar halo velocity anisotropy out to $r \sim 30$ kpc is less radially biased than solar neighborhood measurements. This is opposite to what is expected from violent relaxation, and may indicate the presence of a shell-type structure at $r \sim 24$ kpc. With additional multi-epoch \textit{HST} data, the method presented here has the ability to measure the transverse kinematics of the halo for more stars, and to larger distances. This can yield new improved constraints on the stellar halo formation mechanism, and the mass of the Milky Way.

\end{abstract}

\section{Introduction}

The oldest, and most metal-poor stars in our Galaxy reside in the stellar halo; a diffuse envelope of stars extending out to $r \sim 100$ kpc. The orbital timescales of these halo stars are very long compared to the age of the Galaxy, thus the phase-space structure of the stellar halo is intimately linked to its accretion history. Furthermore, the extreme radial extent of halo stars, well-beyond the baryonic center of our Galaxy, makes them excellent tracers of the dark matter halo.

Stars diffuse more quickly in configuration space as opposed to
angular momentum space, so often the velocity structure of stellar
halo stars provides the strongest link to their initial
conditions. Global kinematic properties, such as the relative pressure
between tangential and radial velocity components, otherwise known as
the velocity anisotropy, can provide important insight into the
formation of the stellar halo. For example, most models of violent
relaxation (e.g. \citealt{diemand04}; \citealt{sales07}), predict an increasingly radially biased velocity
ellipsoid with distance. Local studies, limited to heliocentric
distances $D \lesssim 10$ kpc,
have utilized full three-dimensional (3D) kinematics of halo stars and find a strongly
radially biased velocity anisotropy with $\beta = 1-\langle v_t^2 \rangle/2\langle v_r^2 \rangle \sim 0.5-0.7$ (\citealt{chiba00}; \citealt{gould03}; \citealt{kepley07};\citealt{smith09}; \citealt{bond10}; \citealt{carollo10}). 

Beyond $D \gtrsim 10$ kpc, we are limited to one velocity component:
the line-of-sight (LOS) velocity. At large distances, where $D \gg
R_{0}$, the LOS velocity is almost identical to the radial velocity
component, and hence we have a very poor handle on the tangential
motion of halo stars. Despite this shortcoming, several studies have
used large samples of halo stars, widely distributed over the sky, to
tease out the halo velocity ellipsoid beyond $D \sim 10$
kpc. \cite{sirko04} modeled the LOS velocities of $N \sim 1000$ blue
horizontal branch stars (BHB) selected from the Sloan Digital Sky
Survey (SDSS) data release 4 (DR4) ranging from $10 \lesssim D/\mathrm{kpc}
\lesssim 30$, and found a velocity ellipsoid with $\beta \simeq -0.1
\pm 0.2$, which is consistent with isotropy. More recently, \cite{deason12a} used $N \sim 2000$ BHB stars selected from SDSS DR8 between $20 \lesssim D/\mathrm{kpc} \lesssim 40$, to simultaneously derive the velocity anisotropy and mass profile of the Galaxy; they found a radially biased velocity anisotropy with $\beta \sim 0.5$. Finally, \cite{kafle12} analyzed the SDSS DR8 BHB sample (\cite{xue11}), and found, by dividing the sample into $\sim 15$ radial bins, that the anisotropy profile shows a sharp decline at $r \sim 17$ kpc to $\beta \sim -1$ and is radially biased (with $\beta \sim 0.5$) either side of this apparent dip (see also \citealt{samurovic11}). 

These observations point to a fairly complex velocity anisotropy
profile at large distances, which is likely affected by substructure
in the stellar halo. Unfortunately, analyses using only LOS velocities
require modeling assumptions regarding the underlying potential, and
it is not obvious how such systematics may bias the results. Clearly,
in the ideal case, one would like to \textit{directly measure} the tangential
motions of halo stars. This is a daunting task at large distances in
the halo; at $D \sim 10--100$ kpc, a tangential velocity of $V_t \sim
100$ km s$^{-1}$ corresponds to a proper motion (PM) on the sky of $\mu \sim 2--0.2$ mas yr$^{-1}$. This requires an astrometric accuracy that is un-feasible for current stellar PM surveys. 

\begin{figure*}
  \centering
  \includegraphics[width=16cm, height=6cm]{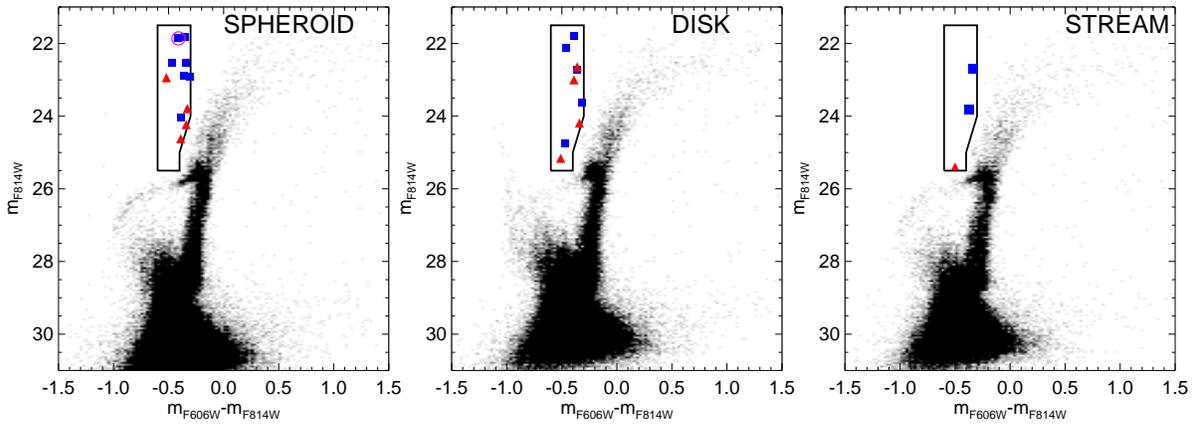}
  \caption[]{\small Color--magnitude diagrams (CMDs) of the three \textit{HST}/ACS
  M31 fields (Spheroid, Disk and Stream), using the photometry from \cite{brown09}. The black box indicates our
  selection of candidate foreground Milky Way halo stars. The symbols
  indicate the stars in each FOV which fall into our CMD selection
  box (11 in Spheroid, 9 in Disk and 3 in Stream). Blue squares are
  Milky Way halo stars, red triangles are M31 stars, and a possible
  Milky Way disk star is highlighted with a magenta circle (see Section \ref{sec:besc}).}
   \label{fig:m31_cmd}
\end{figure*}

The \textit{Hubble Space Telescope} (\textit{HST}) has unparalleled astrometric capabilities;
the displacement of stars over time relative to a stationary
background source can be used to produce very accurate PMs. Quasi-stellar objects (QSOs) have conventionally been used as
reference sources in \textit{HST} PM studies
(e.g. \citealt{kallivayalil06a}; \citealt{kallivayalil06b};
\citealt{piatek08}). In addition, distant background galaxies can also be used
as stationary reference sources. The advantage of background galaxies
is that there are many more of them over the \textit{HST} field-of-view, therefore a
$\sqrt{N}$ averaging advantage means that they can potentially yield
more accurate PM measurements. Galaxy positions can be
difficult to measure (compared with QSOs), but \cite{mahmud08} have
recently developed a template-fitting technique to measure accurate positions of background galaxies. This technique has been
extended by \citep{sohn12a, sohn12b} to measure accurate PMs
of local group galaxies using multi-epoch \textit{HST}
photometry. This work has yielded unprecedented PM accuracy ($\sim
0.01$ mas yr$^{-1}$) and has been used to derive the tangential motion
of M31, located at a distance of $\sim 770$ kpc, and by extension, the
total mass of the local group (\citealt{sohn12a};
\citealt{vandermarel12a}; \citealt{vandermarel12b})

In this study, we make use of the unparalleled PM accuracy achieved by
\cite{sohn12a}, to
extract the PMs of \textit{individual halo stars} in the foreground of
these \textit{HST}/ACS fields. The \textit{HST} fields are small, but deep, which allows
us to study foreground main-sequence (MS) halo stars. This is in contrast
to spectroscopic studies, which cannot go deep enough to study this
dominant stellar halo population. For the first time, we are able to
directly measure the tangential motions of distant, field halo stars. 

The paper is arranged as follows. In Section 2 we describe the multi-epoch \textit{HST}/ACS imaging, and the techniques used to extract PMs of halo stars in these fields. In Section 3 we describe our analysis of the halo stars, and outline our maximum likelihood routine used to determine the halo velocity ellipsoid. Section 4 describes our results, and in Section 5 we discuss the implications of our derived halo velocity anisotropy. Finally, we summarize our main findings in Section 6.

\section{\textit{HST} data}
\begin{table*}
\begin{center}
\renewcommand{\tabcolsep}{0.2cm}
\renewcommand{\arraystretch}{0.75}
\begin{tabular}{c  c  c  c  c  c  c c c}
\hline
Field & RA (J2000) & DEC (J2000) & $m_{\mathrm{F814W}}$ &
  $m_{\mathrm{F606W}}$ & $\mu_l$ [mas yr$^{-1}$] & $\mu_b$ [mas yr$^{-1}$] & Class & Ref\\
\hline
Spheroid & & & &  & & & &\\
\hline
& 00:46:00.25  &  +40:41:37.88   &   24.05  &    23.66 & $-0.09 \pm 0.05$ & $-1.47 \pm 0.06$ & Halo &\\
& 00:46:01.47  &  +40:41:35.53   &   21.86  &    21.45 & $-1.96 \pm
0.04$&  $-2.08 \pm 0.04$& MW Disk? &\\
& 00:46:03.79  &  +40:41:22.81   &   22.53  &    22.19 & $1.36 \pm 0.02$&  $-1.33 \pm 0.02$& Halo &\\
& 00:46:03.67  &  +40:41:56.60   &   22.88  &    22.52 & $2.12 \pm 0.03$&  $-0.82 \pm 0.02$& Halo &\\
& 00:46:04.91  &  +40:41:55.88   &   23.79  &    23.46 & $-0.19 \pm 0.04$& $-0.42 \pm 0.03$ & M31 & V60 (B04)\\
& 00:46:04.92  &  +40:42:47.16   &   24.23  &    23.89 & $0.21 \pm 0.03$&  $-0.28 \pm 0.03$& M31 &\\
& 00:46:06.41  &  +40:42:15.07   &   22.53  &    22.06 & $1.45 \pm 0.02$& $-0.90 \pm 0.02$ & Halo &\\
& 00:46:05.14  &  +40:43:37.19   &   21.82  &    21.47 & $3.91 \pm 0.02$&  $-1.59 \pm 0.02$& Halo &\\
& 00:46:10.05  &  +40:43:06.01   &   22.94  &    22.42 & $0.19 \pm
0.03$&  $+0.22 \pm 0.03$& M31 & V118 (B04; K06)\\
& 00:46:12.92  &  +40:41:22.51   &   22.92  &    22.61 & $1.88 \pm 0.06$&  $-2.83 \pm 0.06$& Halo &\\
& 00:46:12.06  &  +40:42:25.05   &   24.62  &    24.23 & $-0.06 \pm 0.06$& $+0.12 \pm 0.05$ & M31 &\\
\hline
Disk & & & &  & & & &\\
\hline
& 00:49:04.66  &  +42:44:33.36   &   25.16  &    24.65 & $+0.06 \pm 0.11$&  $+0.01 \pm 0.10$& M31 & V5719 (J11)\\
& 00:49:08.91  &  +42:44:13.62   &   21.79  &    21.40 & $-0.59 \pm 0.03$&   $-1.50 \pm 0.04$& Halo &\\
& 00:49:08.30  &  +42:44:50.44   &   22.12  &    21.66 & $+1.03 \pm 0.04$&  $-0.78 \pm 0.04$& Halo &\\
& 00:49:09.77  &  +42:44:51.02   &   24.19  &    23.85 & $-0.31 \pm 0.06$&  $-0.22 \pm 0.07$& M31 & \\
& 00:49:13.50  &  +42:43:36.17   &   22.71  &    22.35 & $-0.71 \pm 0.07$& $-0.67 \pm 0.08$ & Halo &\\
& 00:49:11.84  &  +42:44:25.08   &   23.00  &    22.61 & $-0.20 \pm 0.04$&  $-0.08 \pm 0.04$& M31 &\\
& 00:49:10.54  &  +42:45:25.94   &   22.64  &    22.28 & $+0.16 \pm 0.05$&  $+0.29 \pm 0.04$& M31 & V13779 (J11)\\
& 00:49:13.38  &  +42:45:56.93   &   23.62  &    23.30 & $+2.16 \pm 0.05$&  $-0.40 \pm 0.06$& Halo &\\
& 00:49:13.69  &  +42:45:52.07   &   24.76  &    24.29 & $+0.64 \pm 0.07$& $+0.58 \pm 0.06$ & Halo &\\
\hline
Stream & & & &  & & & &\\
\hline
& 00:44:15.06 &   +39:48:42.50   &   25.38  &    24.88 &$-0.18 \pm 0.08$ &$+0.55 \pm 0.08$ & M31 &\\
& 00:44:26.44 &   +39:47:33.43   &   22.69  &    22.35 &$+0.00 \pm 0.06$ &$-1.85 \pm 0.06$ & Halo &\\
& 00:44:23.93 &   +39:46:26.25   &   23.83  &    23.46 &$-0.43 \pm 0.05$ &$-1.13 \pm 0.07$ & Halo &\\
\hline
\end{tabular}
  \caption{\small The properties of candidate foreground halo stars selected from
  the M31 CMDs. We give the right ascension (RA) and declination (DEC), \textit{HST}/ACS
  STMAG magnitudes, Galactic PMs and designated class (Halo, MW Disk or
  M31 star). The RA, DEC and magnitudes come from \cite{brown09}, and
  the proper motions derive from the study by
  \cite{sohn12a}. Variables stars in M31 identified by
  \citealt{brown04} (B04) and \citealt{jeffery11} (J11) are indicated. One of these stars, V116, was also identified as an M31
  member by \cite{kalirai06} as part of the Spectroscopic and Photometric Landscape of Andromeda's Stellar Halo (SPLASH) survey.}
\label{tab:pms}
\end{center}
\end{table*}

The data employed in our analysis come from three \textit{HST} observing
programs: GO-9453, GO-10265 (PI: T.Brown), and GO-11684 (PI: R.P. van
der Marel).  Programs GO-9453 and
GO-10265 obtained deep optical imaging of three fields in M31
(Spheroid, Disk and Stream: see Figure \ref{fig:m31_cmd}) in two filters
(F606W and F814W) using \textit{HST} ACS/WFC, with
the primary goal of measuring the star formation history in these
fields.  The observations and data reduction are described in
\cite{brown06}, and point source catalogs were distributed in \cite{brown09}.
Program GO-11684 re-observed these fields using \textit{HST} ACS/WFC
and WFC3/UVIS in order to measure the
proper motions of their stars and M31 itself. These data provides five
to seven year baselines with respect to the pre-existing first-epoch data. The second-epoch
observations, data reduction, and determination of proper motions are
described in \cite{sohn12a} (and see below).

\subsection{Proper Motions}

During the course of the \cite{sohn12a} M31 PM study, PM catalogs for \textit{individual} stars in the three \textit{HST} fields were created. As the M31 stars were used to align first- and second-epoch data in the \cite{sohn12a} study, these PMs are measured relative to the M31 stars. To obtain absolute PMs, we added the mean absolute PMs of the M31 stars for each field. For this we used the values reported by \cite{sohn12a}, as determined with respect to stationary background galaxies. The PM errors were calculated based on the positional repeatability seen in the multiple first-epoch ACS/WFC measurements. We adopt similar errors for WFC3/UVIS as for ACS/WFC, which may be a slight overestimate since the WFC3/UVIS pixel scale is smaller.

In the following section we select potential foreground Milky Way halo
stars in the three \textit{HST}/ACS fields. The proper motions of these stars
are then extracted by cross-matching with the \cite{sohn12a} catalog
of individual star PM measurements. In the magnitude range under
consideration ($21.5 \lesssim m_{\mathrm{F814W}} \lesssim 25.5$; see
below), the average uncertainty in the proper motion measurements is
$\sigma_\mu \sim 0.05$ mas yr$^{-1}$. The corresponding velocity error depends on the distance, but for example is only $5$ km s$^{-1}$ at $D = 20$ kpc. This is considerably more
accurate than previous proper motion measurements of individual stars in the halo (with $\sigma_\mu \sim 1--4$ mas yr$^{-1}$; see e.g. \citealt{munn04};
\citealt{casetti06}; \citealt{bramich08})

\subsection{Selection of Foreground Milky Way Halo Stars}
We select candidate foreground Milky Way halo stars using the
color-magnitude diagrams (CMDs) of
the \textit{HST}/ACS M31 fields. The main-sequence turn-off (MSTO) for Milky Way halo stars is located at
brighter (apparent) magnitudes than the M31 MSTO, but at similar
colors blue-ward of $m_{\mathrm{F606W}}-m_{\mathrm{F814W}} \sim
-0.3$. In Figure \ref{fig:m31_cmd}, our color-magnitude selection box is shown on the CMDs of the three M31 fields. We target halo
stars in a sparsely populated region of the CMD, where we expect the
least contamination from Milky Way disk stars and M31 stars. We select 13, 9 and 3 stars from the Spheroid, Disk and Stream fields
respectively, giving a total of 23 candidate halo stars from the three
M31 fields. The extracted PMs and \textit{HST}/ACS STMAG magnitudes are given in Table
\ref{tab:pms}.

\subsection{Comparison with Besan\c{c}on Galaxy model}
\label{sec:besc}
\begin{figure}
  \centering
  \includegraphics[width=8cm, height=6.8cm]{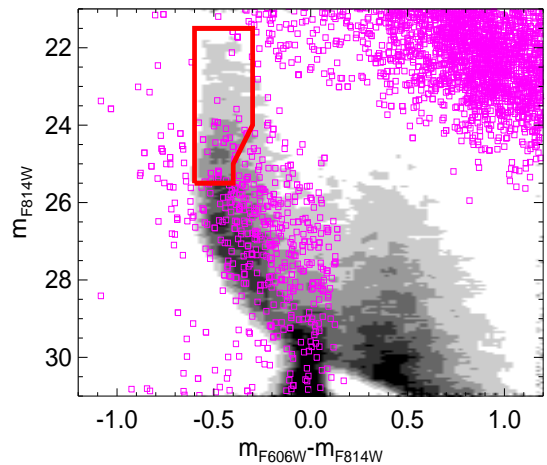}
  \caption[]{\small A CMD for stars in the
  Besan\c{c}on Galaxy model. The FOV is 1 deg$^2$ and is centered on
  M31. The black/gray shaded regions indicate halo stars and the magenta
  squares indicate disk stars (only 25\% of disk stars are shown for clarity). Within our selection box, 7\% of the
  foreground stars are disk stars.}
   \label{fig:besc_cmd}
\end{figure}

\begin{figure}
\centering
\begin{minipage}{\linewidth}
  \includegraphics[width=8cm, height=6.8cm]{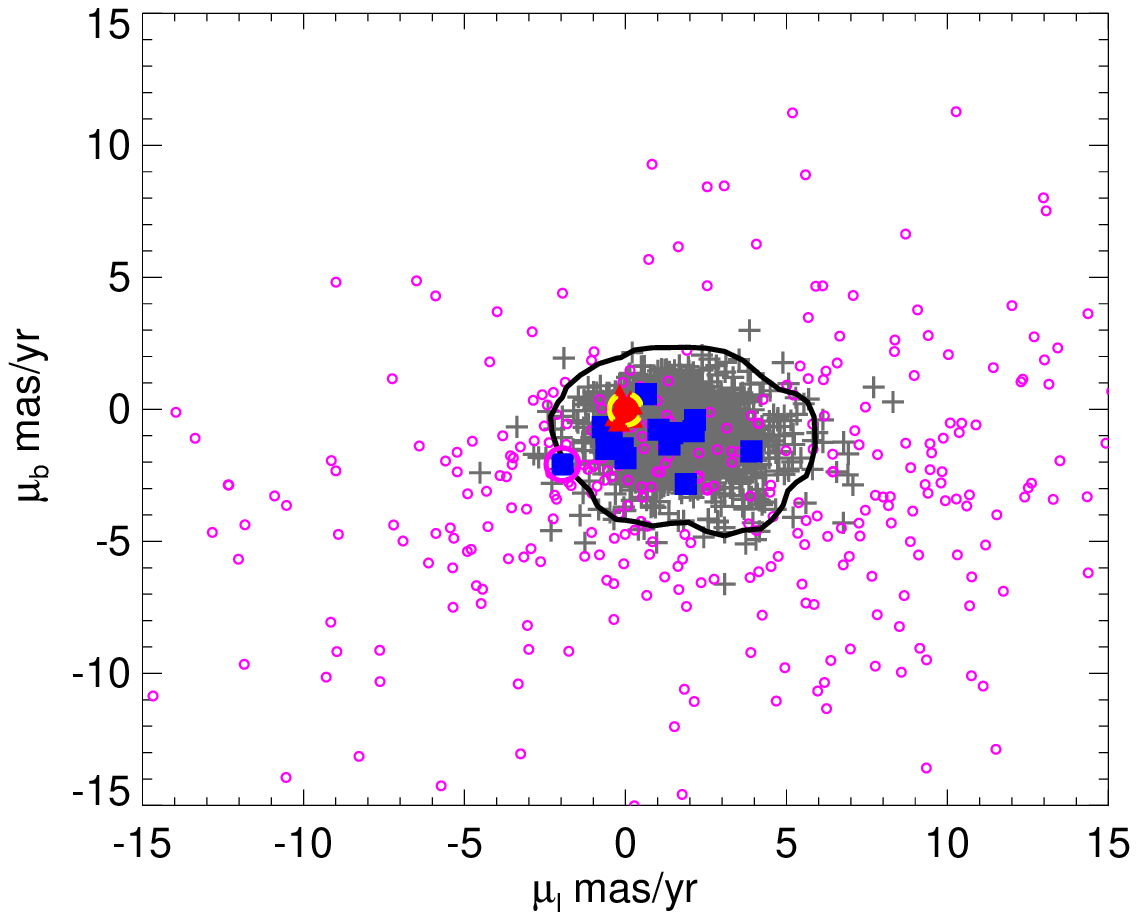}
\end{minipage}
\begin{minipage}{\linewidth}
  \includegraphics[width=8cm, height=6.8cm]{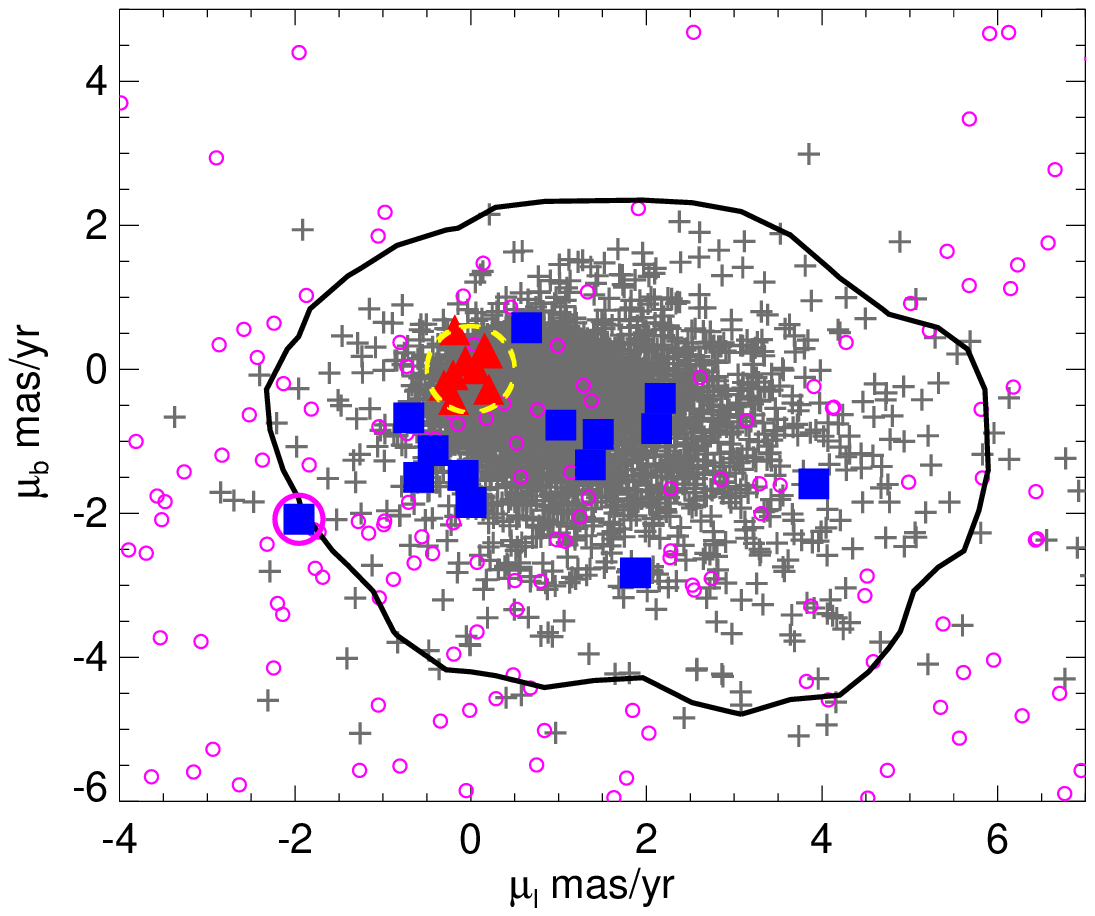}
\end{minipage}
  \caption[]{\small Proper
  motions (PMs) of stars in the Besan\c{c}on model in
  Galactic coordinates ($\mu_l$, $\mu_b$). The stars shown cover a 1
  deg$^2$ FOV centered on M31, and fall in the CMD selection box
  given in Figure \ref{fig:m31_cmd}. The gray crosses are halo stars
  and the magenta circles are disk stars. The blue squares
  show the PMs of stars selected from the M31 CMDs in
  Figure \ref{fig:m31_cmd}. The distribution of PMs is offset in the positive $\mu_l$ direction; this is due to the reflex motion of the Sun. There is a concentration of (9) stars with
  zero PM, consistent with the net PM of M31
  (see \citealt{sohn12a}; \citealt{vandermarel12a}). These stars are
  highlighted with red triangles, and the dashed-yellow circle shows the region
  in PM space consistent with M31 stars. The black contour encompasses
  99\% of the halo population in the Besan\c{c}on model. Outside of this
  contour, the bias caused by including high PM disk stars
  becomes non-negligible. The bottom panel is a zoom-in of the
  top panel.}
   \label{fig:besc_pm}
\end{figure}

In Figure \ref{fig:besc_cmd} we compare our CMD halo
star selection with the Besan\c{c}on Galaxy model (\citealt{besancon}). We
select stars from this model in a 1 deg$^2$ field-of-view (FOV) centered on M31. We
convert Johnson-Cousins \textit{UBVRI} photometry to \textit{HST}/ACS STMAG using the
relations given by \cite{sirianni05}. The stars inside our
color-magnitude selection box are predominantly halo MS stars, and a
small fraction (7 \%) are disk stars. Most of these disk
stars are intrinsically faint white dwarfs, which have similar (blue) colors to MSTO
stars. Note that if we scale this 1 deg$^2$ FOV to the \textit{HST}/ACS FOV then the
Besan\c{c}on model predicts $N \sim 5$ halo stars per \textit{HST}
pointing. Our sample of $N=23$ stars from three pointings is slightly
higher than this prediction; this is due to contamination by M31 stars
in our sample, which we now discuss.

Figure \ref{fig:besc_pm} shows the Galactic proper
motion components of the Besan\c{c}on Galaxy model stars which lie within our color-magnitude selection box. The gray crosses are
halo stars and the magenta circles are disk stars. The disk
stars have a much higher spread in PM than the halo
stars. This is because the stars belonging to the Milky Way disk are
much closer than the halo stars, so they tend to have much larger
proper motions. The black contour indicates the PM region containing
99\% of the halo stars. Within this region there is a very small
amount of disk contamination ($\sim 1\%$), and the low PM disk stars have
a negligible effect on our analysis described in Section \ref{sec:analysis} (see
discussion in Section \ref{sec:disk_contam}). The
blue squares (and red triangles) show the PMs of the 23 stars selected from the
\textit{HST}/ACS images. Encouragingly, the distribution of PMs
closely resembles the halo population of the Besan\c{c}on model. 

The dashed yellow circle indicates the region in PM space consistent with
the net PM of M31 ($(\mu_{l,\mathrm{M}31}, \mu_{b,\mathrm{M}31}) =
(0.04,-0.03) \pm 0.01$ mas yr$^{-1}$;
\citealt{sohn12a}\footnote{\cite{sohn12a} gave values in the (W,N) frame. However, at the location of M31, the (W,N) and (l,b) coordinate frame are aligned to within 2 degrees.}). Note that this region in PM space has a radius of
$\sim 0.5$ mas yr$^{-1}$. There are $N=9$ stars in our sample that
lie within this region, which are highlighted by red triangles. Given the uncertainties in our proper motion measurements and the internal velocity dispersion in the M31 fields ($V_{\rm int} \sim 125$ km s$^{-1}$, see \citealt{vandermarel12a}), we find that these nine stars are typically within 3--4 $\sigma$ of the net motion of M31. In the following
analysis we excise these stars from our sample as they are,
presumably, horizontal branch, asymptotic giant branch and/or variable M31 stars. In fact, four of these stars have been identified as variable stars in M31 from independent studies (\citealt{brown04}; \citealt{jeffery11}). 

 Of the remaining $N=14$ stars, one lies
outside the 99\% contour of the Besan\c{c}on model, this star is
highlighted with a magenta circle. Given the prediction of 7\% disk
contamination from the Besan\c{c}on model, it's likely that we may have
$\sim 1$ disk star in our sample of 14 stars, and the location
of this star in PM space makes it the most probable candidate. In our
analysis below, we only consider stars within the 99\% contour defined
by the Besan\c{c}on model, and so we do not include this star. However, we
note that these two biases in PM space (due to M31 stars and
disk stars) are \textit{fully accounted for} in our
analysis. Furthermore, we verify that a slightly different choice of
contour, which includes this possible disk star, does not change the
final velocity ellipsoid results by more than $\sim 15$ km
s$^{-1}$, which is less than our statistical uncertainties ($\sim 30$ km s$^{-1}$). In the
following section we outline how we model the tangential velocity
components of the remaining $N=13$ halo stars.

\section{Analysis}
\label{sec:analysis}

\subsection{Distance Calibration}
\label{sec:iso}

The tangential motion of the halo stars depends on their PM
and distance: $v_t=4.74047 \, \mu \, D$. Here, the PM is in
mas yr$^{-1}$ and the heliocentric distance is in kpc. 

The large spread in absolute magnitude of stars near the MSTO means
that we cannot determine an accurate distance to individual halo
stars. Therefore, we
cannot measure the velocity ellipsoid directly, despite the fact that we have very accurate PMs. However, we can
determine a probability distribution function (pdf) of halo star distances,
based on a suitably chosen model for the star formation history and 3D
density distribution of the halo (both of which have been well
constrained by existing studies). This allows a statistical
determination of velocity ellipsoid, without knowing the actual 3D
velocities of individual stars.

\begin{figure}
  \centering
  \includegraphics[width=8cm, height=6.8cm]{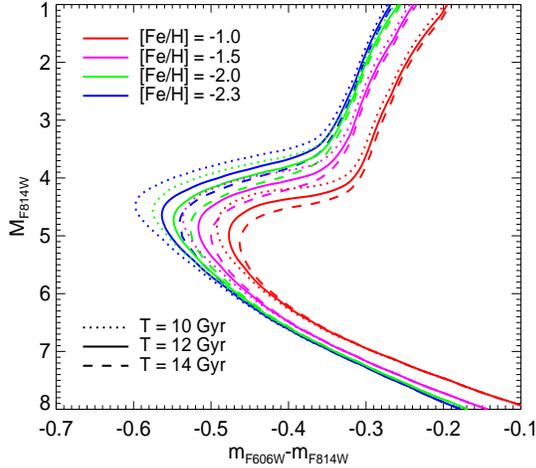}
  \caption[]{\small \cite{vandenberg06} isochrones in the \textit{HST}/ACS
  STMAG system. Metallicities in the range $-2.3 \le [\mathrm{Fe/H}]
  \le -1.0$ are shown by the colored lines, and ages between $10 \le
  T/\mathrm{Gyr} \le 14$ are shown by different line styles. All these
  isochrones have an alpha to iron-peak element ratio $[\alpha/\mathrm{Fe}] =0.3$.}
   \label{fig:isochrones}
\end{figure}
\begin{figure*}
  \centering
  \includegraphics[width=14cm, height=9.3cm]{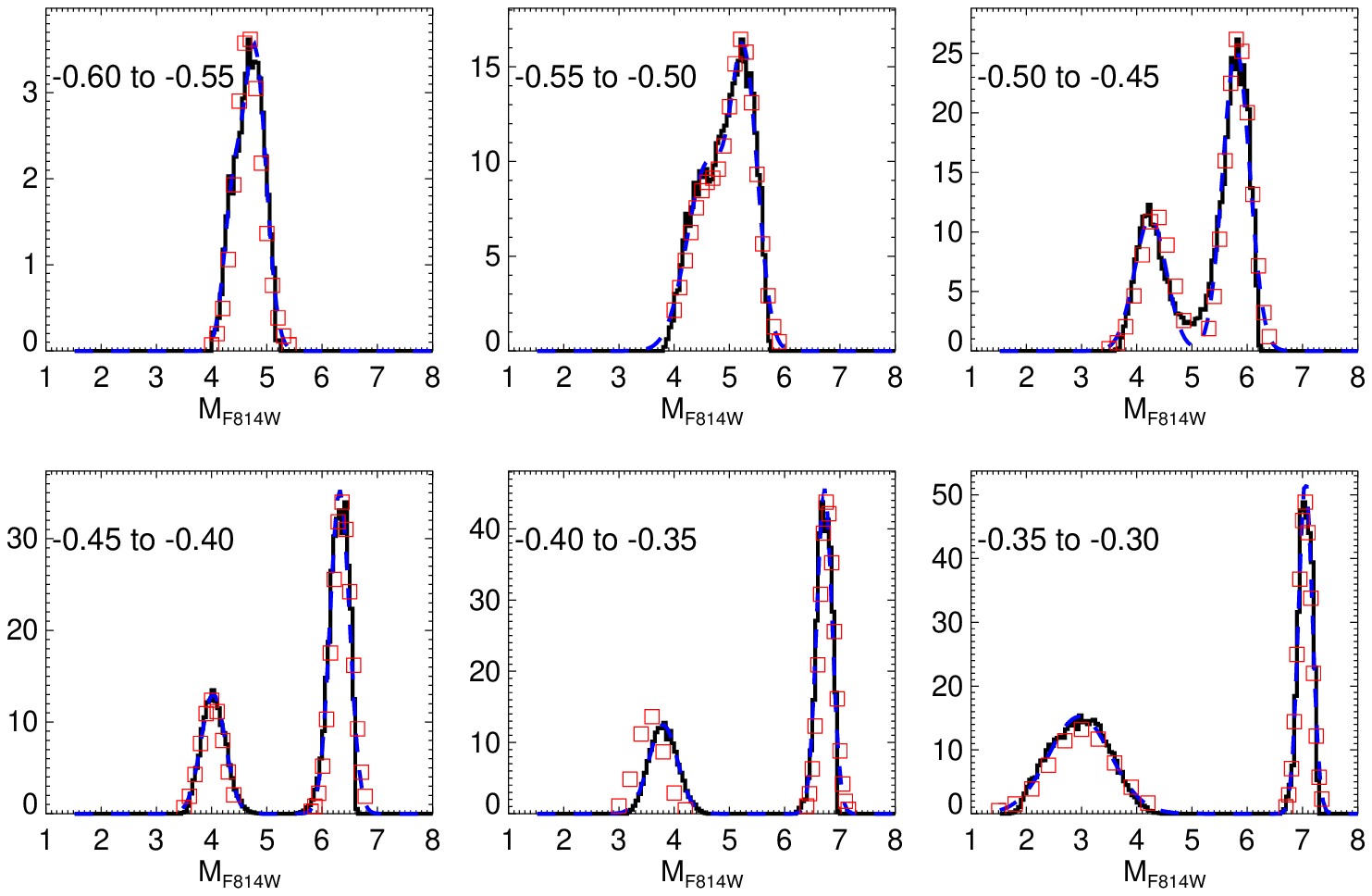}
  \caption[]{\small The $M_{\mathrm{F814W}}$ absolute magnitude
  distribution of the \cite{vandenberg06} isochrones in six color
  bins. The MSTO dominates at bluer colors, while at
  redder colors the MS and RGB populations
  separate into two distinct populations. The blue lines show double-Gaussian fits to the distributions. The red squares indicate the distributions reproduced by our analytic model (see Equation \ref{eq:abm}) in these six color bins.}
   \label{fig:abmag_gauss}
\end{figure*}

We calibrate the pdf of halo star absolute magnitudes via a two step process: First, we use suitably chosen isochrones to describe the distribution of absolute magnitudes as a function of color. Second, we translate this distribution into a continuous, analytic pdf that can be used in a maximum likelihood routine (see Section \ref{sec:ml}).

\subsubsection{Weighted Isochrones}

We use the \cite{vandenberg06} isochrones, calibrated for \textit{HST}/ACS
STMAG photometry (\citealt{brown05}), to model the absolute magnitudes of our selected
halo stars. In Figure \ref{fig:isochrones} we show isochrones for a
range of metallicities ($-2.3 \le [\mathrm{Fe/H}] \le -1.0$) and ages
($10 \le  T/\mathrm{Gyr} \le 14$) applicable for a halo population. In the
  color range of our sample, $m_{\rm F606W}-m_{\rm F814W} < -0.3$, the halo
  stars are close to the MSTO, but they can also lie
  at fainter magnitudes on the MS, or brighter magnitudes on
  the red giant branch (RGB). Thus, the absolute
  magnitudes can range from $3 \lesssim M_{\rm F814W} \lesssim 7.5$, and our sample of halo stars potentially spans a distance range, $10 \lesssim D/\mathrm{kpc}
  \lesssim 100$. 

We model the absolute magnitudes of the halo stars as a function of color, and apply three weighting factors to the isochrones:

\begin{itemize}

\item Initial-mass-function (IMF) weight: We assume a Salpeter IMF and
  weight the contribution from different mass stars accordingly. Low
  mass MS stars thus have higher weight than higher mass
  RGB stars.

\item Metallicity weight: We assume the halo population has a Gaussian
  distribution of metallicities with mean $[\mathrm{Fe/H}] =-1.9$ and
  dispersion $\sigma = 0.5$ (see e.g. \citealt{xue08}). We assume the
  metallicity distribution is constant with radius; this is in good
  agreement with recent studies which find no significant metallicity
  gradient in the stellar halo (Ma Z. et al. in preparation).

\item Age weight: We weight the different age isochrones by assuming a
  Gaussian distribution of ages with mean $\langle T \rangle = 12$ Gyr
  and dispersion $\sigma = 2$ Gyr (see e.g. \citealt{kalirai12}).

\end{itemize}

In Fig. \ref{fig:abmag_gauss} we show the absolute magnitude
distribution in six color bins, weighted by IMF, metallicity and age. The dashed blue line shows a double-Gaussian fit to each color bin. At redder colors the
distribution of absolute magnitudes separates into the MS and RGB populations, while at bluer colors the populations merge into the MSTO. 

\subsubsection{Analytic absolute magnitude calibration}
\begin{figure*}
  \centering
  \includegraphics[width=14cm, height=9.3cm]{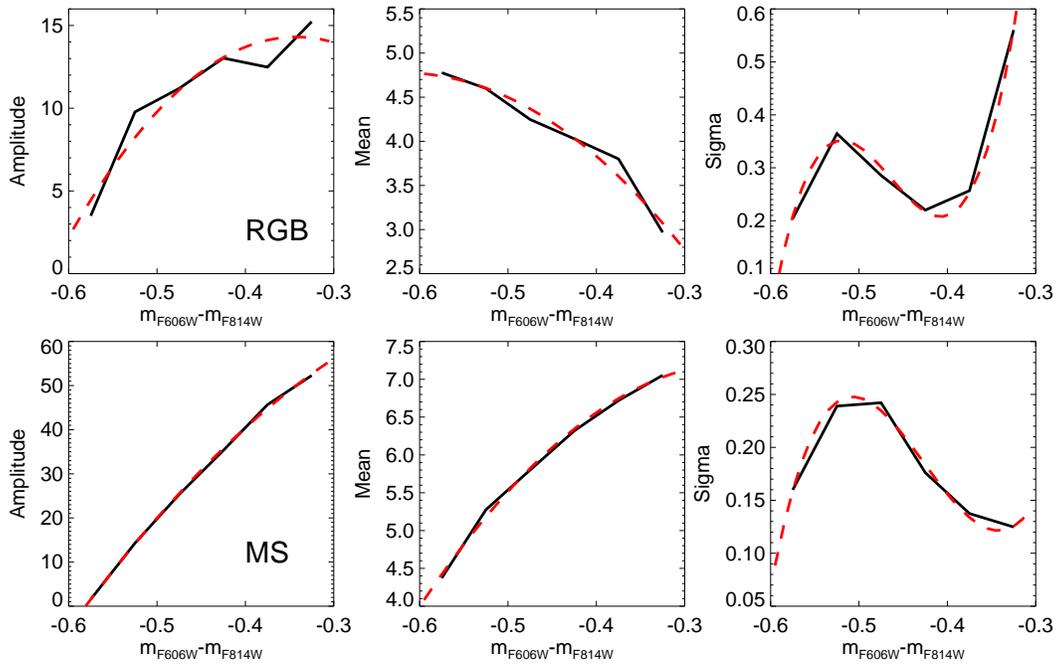}
  \caption[]{\small The Gaussian parameters (amplitude, mean and
  sigma), which define the RGB and MS absolute magnitude calibrations, are shown as a
  function of color. The dashed-red lines show continuous polynomial
  fits (second or third order) to the relations.}
   \label{fig:abmag_params}
\end{figure*}

In Figure \ref{fig:abmag_params} we show the parameters from the double-Gaussian fits (amplitude, mean and sigma) for both the RGB and MS populations as a function of color. The dashed red lines are polynomial (second or third order) fits to the
relations. We use these polynomial relations to define a continuous,
double-Gaussian pdf for absolute magnitude as a function of color:

\begin{eqnarray}
\label{eq:abm}
G(M_{\rm F814W} | m_{\rm F606W} - & m_{\rm F814W})=
G_1(A_1,M_1,\sigma_1,M_{\rm F814W}) \nonumber \\
&+ G_2(A_2, M_2, \sigma_2,M_{\rm F814W})
\end{eqnarray}
where, $G(A, M, \sigma,x)=A \, \mathrm{exp}\left[-\left(x-M\right)^2/(2
  \sigma^2)\right]$, and $A$, $M$ and $\sigma$ (amplitude, mean and sigma) are polynomial functions of
  $m_{\rm F606W}-m_{\rm F814W}$ color.

The red squares in Figure 5. show the absolute magnitude distributions reproduced by our analytic model. These are in good agreement with the (non-analytic) distributions derived from the weighted isochrones.

In the following section, our absolute magnitude pdf is used in a
maximum likelihood routine to determine the tangential components of
the halo velocity ellipsoid.

\subsection{Maximum Likelihood Method}
\label{sec:ml}
In this section, we outline the maximum likelihood method used to
derive the tangential velocity ellipsoid components of the Milky
Way halo. We assume that the velocity distribution in the plane of the
sky follows a Gaussian distribution, with constant ($\sigma_l,
\sigma_b, v_{l,0}, v_{b,0}$) over the radial range spanned by our data:

\begin{equation}
\label{eq:df_vel}
F_v(v_l, v_b) \propto \mathrm{exp}\left[-\frac{\left(v_l-v_{l,0}\right)^2}{2
    \sigma^2_{l}}\right] \mathrm{exp}\left[-\frac{\left(v_b-v_{b,0}\right)^2}{2
    \sigma^2_{b}}\right]
\end{equation}

In the halo, where $D \gg R_0 \simeq 8.5$ kpc, the Galactic coordinates,
$l,b$ are good approximations to the spherical coordinates,
$\phi,\theta$ respectively. For each star, there are six observables: the angular
position on the sky ($l, b$), the PM ($\mu_l, \mu_b$), and
the two photometric magnitudes ($m_{\rm F814W}, m_{\rm F606W}$). Observed
heliocentric velocities are converted to Galactocentric ones by
assuming a circular speed of 240 km s$^{-1}$ (e.g. \citealt{reid09}; \citealt{mcmillan11}; \citealt{schonrich12}) at the position of the
sun ($R_0 = 8.5$ kpc) with a solar peculiar motion ($U, V, W$)=(11.1,
12.24, 7.25) km s$^{-1}$ (\citealt{schonrich10}). Here, $U$ is directed toward the
Galactic center, $V$ is positive in the direction of Galactic rotation
and $W$ is positive towards the North Galactic Pole. In the direction of
M31, the velocity of the sun projects to: $(v_{l}, v_b) =
(-139.5,83.7)$ km s$^{-1}$.

The halo pdf at fixed $m_{\rm F606W}-m_{\rm F814W}$ color, in increments of absolute magnitude, apparent magnitude, Galactic PM and solid angle, $F(y)$, where $y$ is defined as $y=y(M_{\rm F814W}, m_{\rm F814W}, \mu_l, \mu_b, \Omega$), is given by:

\begin{equation}
F \, \, \Delta \textbf{y} \propto F_v \, \rho \, D^5 \, G \, \, \Delta \textbf{y} 
\end{equation}

Here, $D=D(M_{\rm F814W}, m_{\rm F814W})$ is the heliocentric distance, $F_v=F_v(D,\mu_l,\mu_b)$ is the velocity distribution function given in Equation~\ref{eq:df_vel},
$\rho=\rho(D, l,b)$ is the
density distribution of halo stars, $G=G(M_{\rm F814W} | m_{\rm F606W} -
m_{\rm F814W})$ is the absolute magnitude pdf defined in Equation~\ref{eq:abm} and $\Delta
  \textbf{y}= \Delta M_{\rm F814W} \Delta m_{\rm F814W} \Delta \mu_l \Delta \mu_b \Delta \Omega$ is the volume element. We assume the halo stars
follow the broken power-law density profile
derived by \cite{deason11b}, but we comment on the effect of the
density profile parameterization on our results in Section \ref{sec:dens}. We marginalize over the absolute magnitude coordinate,
$\bar{F} = \int F \, \mathrm{d} M_{\rm F814W}$, and define the likelihood
function:

\begin{equation}
\label{eq:like}
L=\prod \bar{F}(\sigma_l, \sigma_b, v_{l,0}, v_{b,0}, \textbf{x})
\end{equation}

Here, $\textbf{x}$ denotes the observables $(m_{\rm F814W},
m_{\rm F606W}-m_{\rm F814W},\mu_l,\mu_b, l, b$), and $\sigma_l, \sigma_b,
v_{l,0}, v_{b,0}$ are free parameters. We use a brute force grid
method to find the maximum likelihood parameters. The net motion in
Galactic latitude, $v_{b,0}$, is set to zero, while we allow for net
motion in Galactic longitude, $v_{l,0} \sim -v_{\phi,0}$, which
approximates the net rotational velocity of the halo. 

The kinematical biases that we have introduced (see
Figure \ref{fig:besc_pm}) are accounted for
through a renormalization of the likelihood function over the region
in observable-space that passes our cuts. In Equation \ref{eq:like},
$\bar{F}(p,\textbf{x})$ is the normalized probability that a star
exists at observables $\textbf{x}$, given model parameters
$p=(\sigma_l, \sigma_b, v_{l,0}, v_{b,0})$. We define $F_{\rm sel}(\textbf{x})$ as
our selection function, which states the probability that a star with
observables $\textbf{x}$ is in our sample. Thus, $F_{\rm sel}(\textbf{x})$ is 1 or
0, depending on our adopted cuts. The normalized probability that a
star is found in our sample, at observables $\textbf{x}$, given model
parameters $p$ is given by: $\bar{F}(p,\textbf{x}) F_{\rm sel}(\textbf{x}) /
N(p)$. Here, $N(p)= \int_{\textbf{x}_c} \bar{F}(p, \textbf{x}) F_{\rm sel}(\textbf{x})$ is the normalization factor, and $\textbf{x}_c$ denotes the
region in $\textbf{x}$-space that passes our selection cuts. We
calculate these integrals numerically. This procedure reduces the
likelihood to:
\begin{equation}
L=\prod \frac{\bar{F}(p,\textbf{x})}{N(p)} 
\end{equation}
since, by definition, $F_{\rm sel}(\textbf{x}_i)= 1$ for any star
  $i=1,...,N$ that makes it into our sample.

In the following Section, we discuss the results of applying this
maximum likelihood algorithm to our sample of halo stars.

\section{Halo velocity anisotropy}
\subsection{Results}
\begin{table*}
\begin{center}
\renewcommand{\tabcolsep}{0.2cm}
\renewcommand{\arraystretch}{2}
\begin{tabular}{|c  c  c  c  c|}
\hline
\multicolumn{5}{|c|}{\textbf{Velocity Ellipsoid [km s$^{-1}$]}} \\
Galactic& $\langle v^2_{\rm los}
\rangle^{1/2}=105^{+5}_{-5}$ &
$\langle v^2_{b} \rangle^{1/2}=83^{+24}_{-16}$ &$\langle
  v^2_{l} \rangle^{1/2}=123^{+29}_{-23}$ & $\langle v_{l}
  \rangle =-75^{+28}_{-29}$  \\
Spherical polars & $\langle v^2_{r}
\rangle^{1/2}=107^{+6}_{-5}$  &
$\langle v^2_{\theta} \rangle^{1/2}=86^{+22}_{-16}$ &$\langle
  v^2_{\phi} \rangle^{1/2}=121^{+28}_{-21}$  & $\langle v_{\phi}
  \rangle =73^{+28}_{-27}$ \\
\hline
\multicolumn{5}{|c|}{\textbf{Velocity Anisotropy}} \\
 & $\beta=0.0^{+0.2}_{-0.4}$ & $\sqrt{\frac{ \langle v^2_t
  \rangle}{ \langle v^2_r \rangle}} =
1.0^{+0.2}_{-0.1}$ & $\sqrt{\frac{\langle v^2_\phi \rangle}{ \langle
  v^2_\theta \rangle}}= 1.4^{+0.4}_{-0.3}$ &\\
\hline
\multicolumn{5}{|c|}{\textbf{Position}} \\
& $l=121^\circ$ & $b=-21^\circ$ & $\langle D \rangle = 19 \pm
1 \pm 6$ kpc& $\langle r \rangle = 24 \pm 1 \pm 6$ kpc \\
\hline
\end{tabular}
  \caption{\small Summary of our main results. We give the velocity ellipsoid in
  Galactic and spherical coordinate systems and the
  resulting velocity anisotropy. Note, the
  LOS velocity dispersion is estimated by previous studies
  in the literature (see text for more details). We also give the
  approximate location of our three \textit{HST} fields in the plane
  of the sky, as well as the average heliocentric and
  Galactocentric distances for our sample. For the latter quantities we list two uncertainties, the first being the error in the mean, and the second being the root-mean-square spread of the sample.}
\label{tab:vtan}
\end{center}
\end{table*}

\begin{figure}
  \centering
  \includegraphics[width=8cm, height=6.8cm]{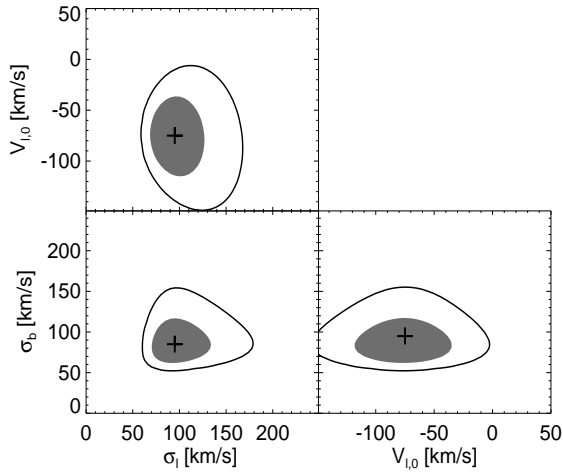}
  \caption[]{\small Likelihood contours for the
  halo velocity ellipsoid in the plane of the sky. The shaded gray regions and solid lines show the 1$\sigma$ and 2$\sigma$ confidence intervals, and the black crosses indicate the maximum likelihood parameters.}
   \label{fig:like}
\end{figure}

In Figure \ref{fig:like} we show the results of our maximum likelihood
analysis (see also Table \ref{tab:vtan}). The resulting velocity
moments\footnote{We also applied our
  likelihood method setting $\langle v_l \rangle \sim - \langle v_\phi
  \rangle =0$. This yields higher $\sigma_\phi$, but the same answer
  for $\sigma^2_\phi+\langle v_\phi \rangle^2$} in the plane of the sky are: $\sigma_b =83^{+24}_{-16}$ km
s$^{-1}$, $\sigma_l=94^{+28}_{-18}$ km s$^{-1}$, and $v_{l,0}=-75^{+28}_{-29}$ km s$^{-1}$. We calculate the
mean and root-mean-square spread of the halo star distances using the pdf of our favored
parameters, e.g. $\langle D \rangle = 1/N \, \sum^N_i \int F_{ml} D
\mathrm{d} D$, where $F_{ml}$ denotes the pdf assuming our maximum likelihood
parameters. We find $\langle D \rangle = 19 \pm 1$ kpc and $\sigma_D = 6
\pm 2$ kpc. In Galactocentric radii, this corresponds to $r \sim 24$ kpc.

Our maximum likelihood velocity ellipsoid quantities are converted
into a spherical polar coordinate system
using a Monte Carlo method. We generate stars which follow the stellar
halo density profile (\citealt{deason11b}), and assign $v_l$ and $v_b$
velocity distributions drawn from our likelihood constraints on
$\sigma_l, \sigma_b$ and $v_{l,0}$. In the approximate distance range
of our sample, $10 \lesssim D/\mathrm{kpc} \lesssim 30$, the
LOS velocity distribution of stellar halo stars, is approximately
Gaussian with $\sigma_{\mathrm{los}} = 105 \pm 5$ km s$^{-1}$ (see
\citealt{sirko04}; \citealt{xue08}; \citealt{brown10}); we apply this LOS velocity distribution to our Monte Carlo
samples. The $v_r, v_\theta, v_\phi$ velocity components are then
calculated from the generated star positions and $v_{\rm los}, v_b,
v_l$ velocities. In Table \ref{tab:vtan} we show the
resulting velocity ellipsoids in spherical polar coordinates. As expected, the $v_\theta$, $v_\phi$ velocity distributions
closely follow the $v_b$, $v_l$ coordinates.

To calculate the velocity anisotropy we consider the ratio of
tangential and radial pressures, i.e. $\langle V^2_t \rangle/ \langle
V^2_r \rangle$, where $\langle V^2_t \rangle = \sigma^2_\phi+\langle
v_\phi \rangle^2 + \sigma^2_\theta$. The resulting velocity anisotropy
suggests isotropic velocity pressure, with $\langle V^2_t \rangle/ \langle
V^2_r \rangle =1.0^{+0.2}_{-0.1}$ or $\beta=0.0^{+0.2}_{-0.4}$, this is in
contrast to the highly radial values found in the solar neighborhood:
$\beta \sim 0.7$ or $(\sigma_r, \sigma_\phi, \sigma_\theta) \sim (140,
80, 80)$ km s$^{-1}$ (\citealt{smith09}; \citealt{bond10}; \citealt{carollo10}). However, several studies estimating velocity
anisotropy from LOS velocity alone, in the distance range, $10 \lesssim
D/\mathrm{kpc} \lesssim 30$, also find a \textit{more
tangential} velocity ellipsoid than the solar neighborhood
(\citealt{sirko04}; \citealt{deason11a}; \citealt{kafle12}). We note
that a significant contribution to the tangential pressure comes from angular
momentum, as $v_\phi \sim 70$ km s$^{-1}$. In the following section,
we discuss some potential systematics which could bias our
results. Encouragingly, we find that systematics play a minor role in
the total error budget.

\subsection{Possible Systematics}
\subsubsection{Stellar Isochrones}
\label{sec:iso_diss}
In Section \ref{sec:iso} we use the \cite{vandenberg06} isochrones to
calibrate our halo star distances. This analysis relies on our
assumptions of metallicity and age for the stellar halo. We adopt
isochrones applicable to an old ($T = 12 \pm 2$ Gyr) and metal-poor
($[\mathrm{Fe/H}]=-1.9 \pm 0.5$) stellar population, but it is worth
remarking how these assumptions may affect our results. Older and/or
more metal-rich populations have fainter absolute magnitudes and hence
higher weight is given to smaller distances in our analysis (and
therefore, lower velocities). For example, the isochrones in the
Besan\c{c}on model assume a similar halo metallicity distribution
($[\mathrm{Fe/H}]=-1.78 \pm 0.5$), but assume an older $14$ Gyr age
population. If we adopt this absolute magnitude calibration in our likelihood
analysis, then our velocity ellipsoid parameters are decreased by ($\Delta
\sigma_l, \Delta \sigma_b) = (-3,-5)$ km s$^{-1}$. Similarly, younger
and/or metal-poorer populations will slightly increase the estimated
velocity dispersions. We find that variations in the mean metallicity
and/or age by 0.5 dex and 2 Gyr respectively, at most can change the
velocity dispersions by $\sim 10$ km s$^{-1}$, which is well below our
observational random errors. We note that more
substantial variations require significant changes in metallicity
and/or age, contrary to observational constraints and theoretical predictions.

\subsubsection{Density Profile}
\label{sec:dens}
We adopt a broken power-law stellar halo density profile, with
halo flattening $q=0.59$, ellipsoidal break radius $r_b=27$ kpc, and
inner and outer power-laws, $\alpha_{\rm in}=2.3$, $\alpha_{\rm
  out}=4.6$ (\citealt{deason11b}). The
uncertainty in the halo star distances means that the density profile
component of the pdf could potentially affect our derived velocity
ellipsoid. For example, a shallower density profile will give higher
weight to larger distances (and hence higher velocities), than a
steeper density profile, and the derived velocity
dispersions will therefore be higher. Encouragingly,
recent studies show general agreement on the form of the stellar
halo density profile out to 50 kpc (see e.g. \citealt{watkins09};
\citealt{sesar11}; \citealt{deason11b}). For example, if we instead adopt the density
profile derived by \cite{sesar11} ($q=0.7$, $r_b=28$ kpc, $\alpha_{\rm
  in}=2.6$, $\alpha_{\rm  out}=3.8$), our velocity ellipsoid values
are increased by ($\Delta \sigma_l, \Delta \sigma_b) = (6,3)$ km
s$^{-1}$. These are very small differences relative to the statistical
uncertainties in our results ($\sim 30$ km s$^{-1}$). Note that a very
steep halo density profile ($\alpha \gtrsim 6$) is required to
significantly \textit{decrease} our tangential velocity ellipsoid values.
 
We also note that the systematics primarily associated with distances, affect our estimate of $ \langle V^2_t \rangle$
more than the ellipsoid ratio $ \langle V^2_\phi \rangle / \langle V^2_\theta
\rangle$. This is because any systematic in our distance calibration,
affects both tangential components by a similar degree.

\subsubsection{Disk Contamination}
\label{sec:disk_contam}
The Besan\c{c}on Galaxy model predicts that the majority of stars selected from
our color-magnitude cut are halo stars, but $\sim 7\%$ of the stars
could belong to the Milky Way disk. Our further restriction in proper
motion space (see Figure \ref{fig:besc_pm}), reduces this level of
contamination even further (to 1\% in the Besan\c{c}on model). Even so, we
have quantified any bias due to the remaining disk stars. We select $N=13$ stars at random from the Besan\c{c}on
model using our color-magnitude and PM selection, and apply
our maximum likelihood analysis. We repeat this process using $N=13$
\textit{pure halo stars} from the Besan\c{c}on model. After $10^4$ trials,
we can quantify any biases due to contamination by the small number of disk stars. We
find that any biases are negligible ($< 5$ km s$^{-1}$), and can be
safely ignored in the current analysis. However, we note that if we apply a
less restrictive contour in proper motion space (see
Figure \ref{fig:besc_pm}), then these biases can increase. 

\subsubsection{Solar Motion}

In recent years, the solar motion --- and in particular, the azimuthal
velocity of the Sun --- has come under renewed scrutiny. There has
been some debate over whether the circular velocity at the position of
the Sun ($R_0=8.5$ kpc) is close to the IAU-recommended value, $V_c = 220$ km
s$^{-1}$, or if it needs to be revised upwards to $V_c = 240$ km
s$^{-1}$ (e.g. \citealt{reid09}; \citealt{mcmillan11};
\citealt{schonrich12}). Recently, \cite{bovy12} showed that the
azimuthal velocity of the sun is close to $v_\phi \sim 240$ km
s$^{-1}$. However, the authors claim that this is due to an offset
between the azimuthal velocity of the Local Standard of Rest and the local circular
velocity, rather than a change in the circular velocity. In either case, the net solar motion --- the quantity of importance in this study --- remains uncertain.

In the context of the present study, variation of the azimuthal solar motion by $\sim 20$
km s$^{-1}$ can change the projection of the solar motion in the
direction of M31 by approximately ($|\Delta v_l|, \Delta v_b|$) $\sim
(10,5)$ km s$^{-1}$. This will most notably affect our estimate of net
rotation in the halo by $\sim 10$ km s$^{-1}$. Thus, even with a
decrease in the Sun's rotational velocity of 20 km s$^{-1}$, we still find a significant
signal of rotation in our sample. We note that the velocity anisotropy
is largely unaffected by this translation.

 \begin{figure*}
  \centering
  \includegraphics[width=15cm, height=10cm]{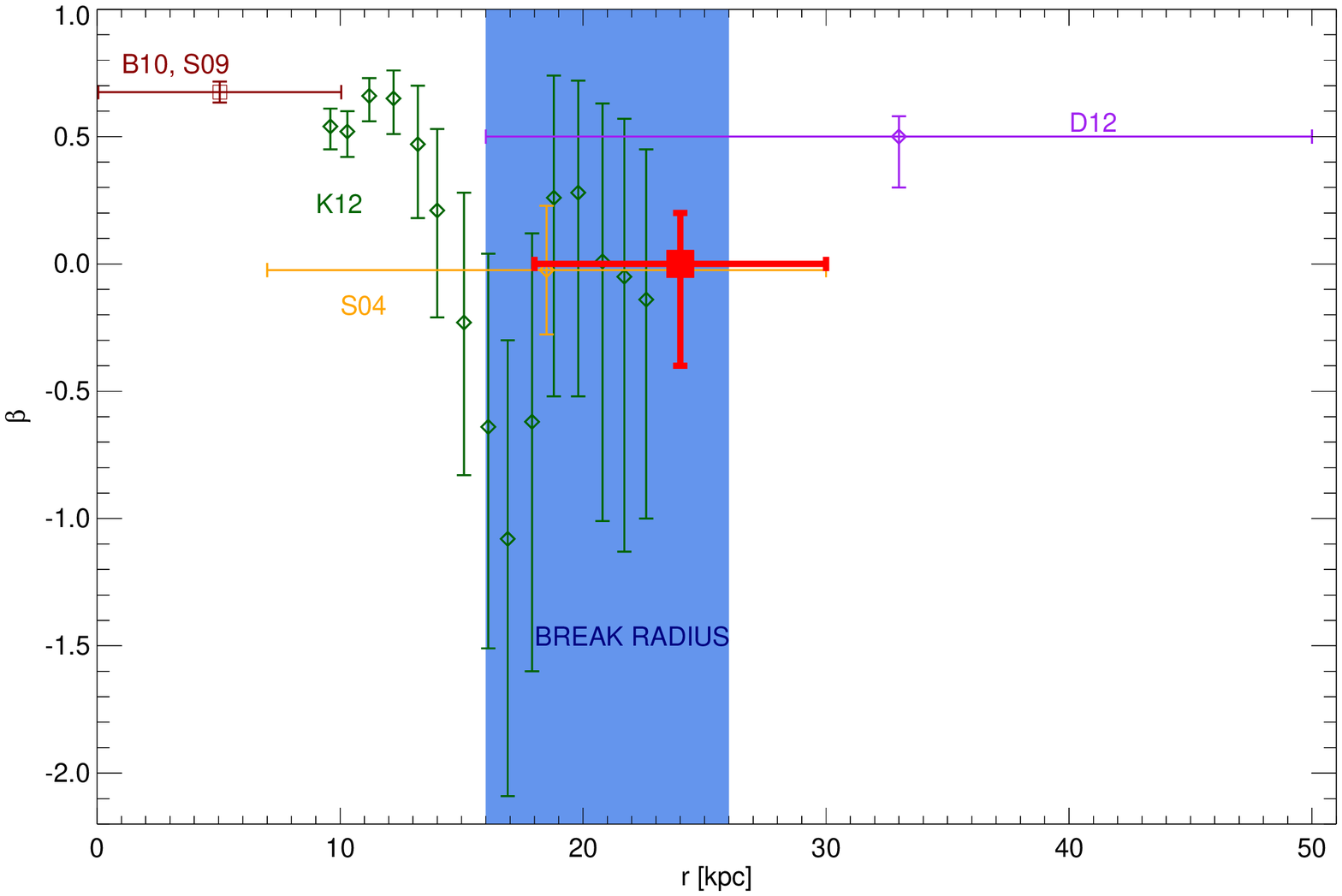}
  \caption[]{\small Velocity anisotropy as a of function radius. The
  colored error bars show the estimates from nearby halo stars with
  full 3D velocity measurements (\citealt{bond10}: B10; \citealt{smith09}: S09), and
  distant ($D \gtrsim 10$ kpc) halo stars with LOS velocity
  measurements (\citealt{sirko04}: S04; \citealt{deason12a}: D12;
  \citealt{kafle12}: K12). The
  red error bar shows our measure from $N=13$ halo stars with
  PM measurements in the radial range $18 \lesssim
  r/\mathrm{kpc} \lesssim 30$. The blue shaded region indicates the
  approximate radial range of the `break radius' in the Milky Way
  stellar halo ($16 \lesssim r/\mathrm{kpc} \lesssim 26$: \citealt{deason11b})}.
   \label{fig:beta}
\end{figure*}

\section{Discussion}

Our measurement of an \textit{isotropic} velocity anisotropy at $r
\sim 24$ kpc, is at odds with theoretical predictions of violent
relaxation, where $\beta$ tends to \textit{increase} with
radius: Our result of $\beta = 0.0^{+0.2}_{-0.4}$ is lower than solar
neighborhood measurements by $\sim 3\sigma$. In this section, we discuss some possible explanations for
this surprising result.

\subsection{Global Substructure: A Galactic Shell?}
\label{ref:shell}
In Figure \ref{fig:beta} we show velocity anisotropy as a function of
radius, based on various estimates from the literature. When our
results are compared to solar neighborhood measures of $\beta$ at smaller
radii, and LOS velocity estimates at larger radii, we
begin to see that the velocity anisotropy may have a `dip' at $r \sim
20$ kpc, rather than a constant or continuous decline. Furthermore, the location
of this possible dip is coincident with a break in the stellar halo
density: \cite{deason11b} (see also \citealt{sesar11};
\citealt{watkins09}) find that the stellar halo density profile
declines more rapidly beyond a break radius of $r \sim 16-26$ kpc. The
origin of this break radius is still uncertain: \cite{deason13}
recently suggested that the break radius in the Milky Way may be due to a
shell-type structure built up from the aggregation of accreted stars at
apocenter. However, \cite{beers12} claim that the break radius
signifies a transition between two halo populations of different
origin (see discussion below). In the former scenario, the kinematic
signature of a shell --- a predominance of tangential motion at the
turn-around radius, plus a cold radial velocity dispersion --- is
intriguingly similar to the apparent `dip' in velocity anisotropy that
we observe. At present, the evidence for a rise again in velocity anisotropy
beyond $r \sim 30$ kpc, relies solely on LOS velocities, where the derived anisotropy is intimately linked to the assumptions of the halo
potential. Therefore, it is important to independently determine the tangential
motion of halo stars $\textit{beyond}$ the break radius in order to
confirm/falsify the apparent dip in anisotropy.

It is possible that some of our halo stars belong to the ``TriAnd"
overdensity. This structure subtends an area of at
least $50^\circ \times 60^\circ$ in the constellations of Triangulum
and Andromeda, and is located at a similar distance to our
halo sample ($16 \lesssim D/\mathrm{kpc}  \lesssim 25$, \citealt{rocha04}; \citealt{majewski04}; \citealt{martin07}). The nature of TriAnd has been debated in the literature; \cite{rocha04} (see also \citealt{majewski04}) propose either a bound core of a very dark matter dominated dwarf or a portion of a tidal stream, while Johnston et al. (2012) suggest that TriAnd is an apogalactian piece of a disrupted dwarf galaxy. The conclusions we draw here are most consistent with this last interpretation.

\subsection{Stellar Halo Formation Mechanism: In-Situ Stars?}
In recent years, an additional formation mechanism for halo stars has
been put forward. Studies based on hydrodynamical cosmological
simulations suggests that halo stars can form \textit{in-situ} from
gas in the parent galaxy in addition to formation in external
dwarf galaxies and then subsequent accretion (\citealt{zolotov09};
\citealt{font11}). \cite{mccarthy12} showed that these
in-situ stars can have significant prograde rotation and
therefore increased tangential pressure support from angular
momentum. Therefore, this alternative formation mechanism for halo
stars could also explain the more tangentially biased halo star
orbits. However, this does not explain why the orbits of halo stars in the
solar neighborhood --- presumably where \textit{in-situ} stars are more
dominant over accreted stars --- have such strongly radial orbits.

\subsection{A Cold Stream?}
Finally, our sample of halo stars is small ($N =13$) and covers a
narrow FOV, so we cannot reject the possibility that our results are
biased by local substructure. However, we do not find any obvious
clustering of our sample in PM space, so
it seems unlikely that the majority of our halo stars belong to a common
stream. Furthermore, the agreement of our results with studies using
LOS velocities in a similar radial regime but different directions on
the sky,  suggests that the
isotropic velocity anisotropy derived here is a true global property.

\subsection{Halo Rotation}
We find a significant signal of prograde rotation in our halo sample,
with $\langle v_\phi \rangle \sim 70$ km s$^{-1}$. As mentioned above,
this signal could be due to the presence of a shell-type structure in
this radial regime, or the halo could have a net rotation --- perhaps due
to the influence of stars born \textit{in-situ}. Previous work on the
rotation of halo populations have found contrasting
results. \cite{frenk80} and \cite{zinn85} found evidence of net
prograde rotation in the Milky Way globular cluster
population (with $\langle v_\phi \rangle \simeq 50--60$ km
s$^{-1}$). More recently, \cite{deason11a} find that BHB stars with
$[\mathrm{Fe/H}] > -2$ have a net streaming motion of $\langle v_\phi
\rangle \sim 50$ km s$^{-1}$ in a similar radial range to this study. However,
\cite{carollo10} claim that their sample of outer ($D > 15$ kpc) halo
stars have a net \textit{retrograde} rotation with $v_\phi \sim -80$
km s$^{-1}$ (but see \citealt{schonrich11}). 

It is clear that a confused picture surrounds the rotational
properties of the halo. This is something we hope to
address in the future with additional \textit{HST} fields.

\subsection{Future Work}
The above discussion illustrates the importance of measuring the
radial velocity anisotropy profile in order to understand the
formation mechanism of the stellar halo. We hope to extend this
present study to multiple \textit{HST} fields with long-baseline multi-epoch
photometry. This will enable us to move beyond small number
statistics, cover a wider area on the sky, and probe further out into
the stellar halo. Studies using LOS velocity measurements rely on
assumptions of the halo potential and are limited to within $r
\lesssim 30-40$ kpc: using PMs, we can more directly measure the
tangential motion of halo stars. Furthermore, the depth of the \textit{HST}
fields allows us to study MS stars; the dominant population
of the stellar halo. Current spectroscopic studies and the upcoming
\textit{Gaia} mission (with magnitude limit $V < 20$) are limited to
intrinsically bright halo stars, such as BHB or RGB stars, which may
not be unbiased tracers.

Finally, an \textit{independent} measure of velocity anisotropy is
vital in order to derive the mass profile of our
Galaxy. Recently, \cite{deason12b} found that the radial velocity
dispersion of halo stars declines rapidly at large radii. At present,
the degeneracy between tracer density, anisotropy and halo mass cannot
be disentangled. However, a measure of the tangential motion of these
distant halo stars will allow us to address whether or not this cold radial
velocity dispersion is due to a shift in pressure from radial to
tangential components. Our method, using multi-epoch \textit{HST} images, may
be the \textit{only} way to measure velocity anisotropy at these large
distances. The upcoming \textit{Gaia} mission will measure proper
motions for an unprecedented number of halo stars with $V < 20$, and
will likely revolutionize our understanding of the inner stellar
halo. However, even with bright halo tracers (e.g. BHB or Carbon stars), the
PM accuracy of \textit{Gaia}, $\sigma_\mu \sim 0.3$ mas yr$^{-1}$ ($\sigma_V \sim
140$ km s$^{-1}$ at $D \sim 100$ kpc), will be unable to accurately
constrain the tangential motion of very distant halo stars. 

The PM accuracies achievable with just a few orbits of \textit{HST} time (see
Table \ref{tab:pms}), are well below the halo velocity dispersion,
even at $D=100$ kpc. Therefore, PM accuracy is no impediment at all
for studies of this kind. However, the small FOV of \textit{HST} means
that there are only a handful of halo stars per field. Therefore, to
find the rarer stars that are at larger distances than the mean for
our sample ($\langle r \rangle \sim 24$ kpc), \textit{many} fields
need to be observed.

\section{Conclusions}

We derive proper motions (PMs) for $N=13$ Milky Way halo stars from three
\textit{HST}/ACS fields with long baseline (5-7 years), multi-epoch
photometry. The unprecedented accuracy of these PM
measurements ($\sigma_\mu \sim 0.05$ mas yr$^{-1}$) allows us to derive the
tangential velocity ellipsoid of halo stars between $18 \lesssim
r/\mathrm{kpc} \lesssim 30$. Until now, measurements of the tangential motion
of halo stars have been restricted to within $D < 10$
kpc. This study is the first step towards measuring the halo velocity
ellipsoid of the stellar halo beyond the solar neighborhood,
independently of any assumptions regarding the underlying halo
potential. We summarize our conclusions as follows:

\medskip
(1) We select $N=23$ candidate halo stars from three \textit{HST}/ACS
    fields centered on M31. These stars are selected from the halo MSTO
    region of the CMD. The individual PMs are extracted
    from the multi-epoch \textit{HST} photometry. Inspection of the proper
    motions of these stars shows that nine belong to M31, and one is a
    possible Milky Way disk star. The selected halo stars have magnitudes
    ranging from $21.5 < m_{\rm F814W} < 24.5$ and are MS and/or RGB stars near the MSTO. The distances to these stars are calibrated from stellar
    isochrones applicable to an old, metal-poor halo population.

\medskip
(2) We devise a maximum likelihood routine to derive the tangential
    velocity ellipsoid of these halo stars. We measure PMs for
    individual stars, but not their distance or radial velocities. Therefore, our
    determination of the velocity ellipsoid is statistical in nature. We find velocity
    dispersions $\sigma_b = 83^{+24}_{-16}$ km s$^{-1}$, $\sigma_l =
    94^{+28}_{-18}$ km s$^{-1}$ and mean streaming motion,
    $v_{l,0}=-75^{+28}_{-29}$ km s$^{-1}$. In the distance range of the halo star
    sample, $18 \lesssim r/\mathrm{kpc} \lesssim 30$, the Galactic
    velocity components closely approximate the spherical
    coordinate system ($v_\phi \sim -v_l$ and $v_\theta \sim v_b$).

\medskip
(3) The tangential velocity components are combined with independent
    measures of the radial velocity dispersion ($\sigma_{r} \sim
    \sigma_{\rm los} = 105 \pm 5$ km s$^{-1}$) in this radial regime to estimate the
    velocity anisotropy of the stellar halo. We find approximate isotropy between
    radial and tangential velocity second moments with $\beta = 0.0^{+0.2}_{-0.4}$. This
    is in contrast to the strongly radial anisotropy of halo stars found in the
    solar neighborhood ($\beta \sim 0.7$). The increased tangential
    relative to radial pressure has a significant contribution from
    angular momentum as $\langle v_\phi \rangle \sim 70$ km
    s$^{-1}$. We note that the more tangentially biased velocity
    anisotropy outside of the solar neighborhood is mainly due to a
    \textit{decrease} in radial pressure (from $140$ to
    $100$ km s$^{-1}$), rather than a significant increase in
    tangential pressure.
      
\medskip
(4) The radial anisotropy profile is poorly constrained, especially at
    large radii. However,
    there is growing evidence that there may be an isotropic/tangential
    `dip' in velocity
    anisotropy in the radial range $15 \lesssim r/\mathrm{kpc}
    \lesssim 25$. Intriguingly, this coincides with a break in the
    stellar density profile, beyond which the stellar density
    falls off more rapidly. These two lines of
    evidence suggest that there may be a shell-type structure in this
    radial regime. However, at present we cannot discount the
    influence of halo stars formed \textit{in-situ} which may have
    more tangentially biased orbits than accreted stars.

\section*{Acknowledgments}
We thank Jay Anderson for his collaboration on the proper motion analysis reported in \cite{sohn12a}, and we thank an anonymous referee for a thorough and constructive report. AJD thanks Prajwal Kafle for kindly providing his data for Figure 8. Support for this work was provided by NASA through a grant for program GO-11684 from the Space Telescope Science Institute (STScI), which is operated by the Association of Universities for Research in Astronomy
(AURA), Inc., under NASA contract NAS5-26555. AJD is currently supported by NASA through Hubble Fellowship grant HST-HF-51302.01, awarded by the Space Telescope Science Institute, which is operated by the Association of Universities for Research in Astronomy, Inc., for NASA, under contract NAS5-26555.

\textit{Facility:} \textit{HST} (ACS/WFC; WFC3/UVIS)
\label{lastpage}
\bibliography{mybib}

\begin{thebibliography}{60}
\expandafter\ifx\csname natexlab\endcsname\relax\def\natexlab#1{#1}\fi

\bibitem[{Beers {et~al.}(2012)}]{beers12}
Beers, T.~C., {et~al.} 2012, \apj, 746, 34

\bibitem[{Bond {et~al.}(2010)}]{bond10}
Bond, N.~A., {et~al.} 2010, \apj, 716, 1

\bibitem[{Bovy {et~al.}(2012)}]{bovy12}
Bovy, J., {et~al.} 2012, \apj, 759, 131

\bibitem[{Bramich {et~al.}(2008)}]{bramich08}
Bramich, D.~M., {et~al.} 2008, \mnras, 386, 887

\bibitem[{{Brown} {et~al.}(2004){Brown}, {Ferguson}, {Smith}, {Kimble},
  {Sweigart}, {Renzini}, \& {Rich}}]{brown04}
{Brown}, T.~M., {Ferguson}, H.~C., {Smith}, E., {Kimble}, R.~A., {Sweigart},
  A.~V., {Renzini}, A., \& {Rich}, R.~M. 2004, \aj, 127, 2738

\bibitem[{Brown {et~al.}(2005)}]{brown05}
Brown, T.~M., {et~al.} 2005, \aj, 130, 1693

\bibitem[{Brown {et~al.}(2006)}]{brown06}
---. 2006, \apj, 652, 323

\bibitem[{Brown {et~al.}(2009)}]{brown09}
---. 2009, \apjs, 184, 152

\bibitem[{{Brown} {et~al.}(2010){Brown}, {Geller}, {Kenyon}, \&
  {Diaferio}}]{brown10}
{Brown}, W.~R., {Geller}, M.~J., {Kenyon}, S.~J., \& {Diaferio}, A. 2010, \aj,
  139, 59

\bibitem[{Carollo {et~al.}(2010)}]{carollo10}
Carollo, D., {et~al.} 2010, \apj, 712, 692

\bibitem[{{Casetti-Dinescu} {et~al.}(2006){Casetti-Dinescu}, {Majewski},
  {Girard}, {Carlin}, {van Altena}, {Patterson}, \& {Law}}]{casetti06}
{Casetti-Dinescu}, D.~I., {Majewski}, S.~R., {Girard}, T.~M., {Carlin}, J.~L.,
  {van Altena}, W.~F., {Patterson}, R.~J., \& {Law}, D.~R. 2006, \aj, 132, 2082

\bibitem[{{Chiba} \& {Beers}(2000)}]{chiba00}
{Chiba}, M., \& {Beers}, T.~C. 2000, \aj, 119, 2843

\bibitem[{{Deason} {et~al.}(2011a){Deason}, {Belokurov}, \&
  {Evans}}]{deason11a}
{Deason}, A.~J., {Belokurov}, V., \& {Evans}, N.~W. 2011a, \mnras, 411, 1480

\bibitem[{{Deason} {et~al.}(2011b){Deason}, {Belokurov}, \&
  {Evans}}]{deason11b}
---. 2011b, \mnras, 416, 2903

\bibitem[{{Deason} {et~al.}(2012{\natexlab{a}}){Deason}, {Belokurov}, {Evans},
  \& {An}}]{deason12a}
{Deason}, A.~J., {Belokurov}, V., {Evans}, N.~W., \& {An}, J.
  2012{\natexlab{a}}, \mnras, 424, L44

\bibitem[{{Deason} {et~al.}(2012{\natexlab{b}}){Deason}, {Belokurov}, {Evans},
  {Koposov}, {Cooke}, {Pe{\~n}arrubia}, {Laporte}, {Fellhauer}, {Walker}, \&
  {Olszewski}}]{deason12b}
{Deason}, A.~J., {et~al.} 2012{\natexlab{b}}, \mnras, 425, 2840

\bibitem[{{Deason} {et~al.}(2013){Deason}, {Belokurov}, {Evans},
  \& {Johnston}}]{deason13}
{Deason}, A.~J., {Belokurov}, V., {Evans}, N.~W., \& {Johnston}, K.~V.
  2013, ApJ, 763, 113

\bibitem[{{Diemand} {et~al.}(2004){Diemand}, {Moore}, \& {Stadel}}]{diemand04}
{Diemand}, J., {Moore}, B., \& {Stadel}, J. 2004, \mnras, 352, 535

\bibitem[{{Font} {et~al.}(2011){Font}, {McCarthy}, {Crain}, {Theuns}, {Schaye},
  {Wiersma}, \& {Dalla Vecchia}}]{font11}
{Font}, A.~S., {McCarthy}, I.~G., {Crain}, R.~A., {Theuns}, T., {Schaye}, J.,
  {Wiersma}, R.~P.~C., \& {Dalla Vecchia}, C. 2011, \mnras, 416, 2802

\bibitem[{{Frenk} \& {White}(1980)}]{frenk80}
{Frenk}, C.~S., \& {White}, S.~D.~M. 1980, \mnras, 193, 295

\bibitem[{{Gould}(2003)}]{gould03}
{Gould}, A. 2003, \apj, 583, 765

\bibitem[{Jeffery {et~al.}(2011)}]{jeffery11}
Jeffery, E.~J., {et~al.} 2011, \aj, 141, 171

\bibitem[{Johnston {et~al.}(2012)}]{johnston12}
Johnston, K. V., Sheffield, A. A., Majewski, S. R., Sharma, S., \& Rocha-Pinto, H. J. 2012, \apj, 760, 95

\bibitem[{{Kafle} {et~al.}(2012){Kafle}, {Sharma}, {Lewis}, \&
  {Bland-Hawthorn}}]{kafle12}
{Kafle}, P.~R., {Sharma}, S., {Lewis}, G.~F., \& {Bland-Hawthorn}, J. 2012, \apj, 761, 98

\bibitem[{{Kalirai}(2012)}]{kalirai12}
{Kalirai}, J.~S. 2012, \nat, 486, 90

\bibitem[{{Kalirai} {et~al.}(2006){Kalirai}, {Guhathakurta}, {Gilbert},
  {Reitzel}, {Majewski}, {Rich}, \& {Cooper}}]{kalirai06}
{Kalirai}, J.~S., {Guhathakurta}, P., {Gilbert}, K.~M., {Reitzel}, D.~B.,
  {Majewski}, S.~R., {Rich}, R.~M., \& {Cooper}, M.~C. 2006, \apj, 641, 268

\bibitem[{{Kallivayalil} {et~al.}(2006b){Kallivayalil}, {van der Marel}, \&
  {Alcock}}]{kallivayalil06b}
{Kallivayalil}, N., {van der Marel}, R.~P., \& {Alcock}, C. 2006b, \apj, 652,
  1213

\bibitem[{{Kallivayalil} {et~al.}(2006a){Kallivayalil}, {van der Marel},
  {Alcock}, {Axelrod}, {Cook}, {Drake}, \& {Geha}}]{kallivayalil06a}
{Kallivayalil}, N., {van der Marel}, R.~P., {Alcock}, C., {Axelrod}, T.,
  {Cook}, K.~H., {Drake}, A.~J., \& {Geha}, M. 2006a, \apj, 638, 772

\bibitem[{Kepley {et~al.}(2007)}]{kepley07}
Kepley, A.~A., {et~al.} 2007, \aj, 134, 1579

\bibitem[{{Mahmud} \& {Anderson}(2008)}]{mahmud08}
{Mahmud}, N., \& {Anderson}, J. 2008, \pasp, 120, 907

\bibitem[{Majewski {et~al.}(2004)}]{majewski04}
Majewski, S. R., Ostheimer, J. C., Rocha-Pinto, H. J., et al. 2004, \apj, 615, 738

\bibitem[{Martin {et~al.}(2007)}]{martin07}
Martin, N. F., Ibata, R. A., \& Irwin, M. 2007, \apj, 668, L123

\bibitem[{{McCarthy} {et~al.}(2012){McCarthy}, {Font}, {Crain}, {Deason},
  {Schaye}, \& {Theuns}}]{mccarthy12}
{McCarthy}, I.~G., {Font}, A.~S., {Crain}, R.~A., {Deason}, A.~J., {Schaye},
  J., \& {Theuns}, T. 2012, \mnras, 420, 2245

\bibitem[{{McMillan}(2011)}]{mcmillan11}
{McMillan}, P.~J. 2011, \mnras, 414, 2446

\bibitem[{Munn {et~al.}(2004)}]{munn04}
Munn, J.~A., {et~al.} 2004, \aj, 127, 3034

\bibitem[{{Piatek} {et~al.}(2008){Piatek}, {Pryor}, \& {Olszewski}}]{piatek08}
{Piatek}, S., {Pryor}, C., \& {Olszewski}, E.~W. 2008, \aj, 135, 1024

\bibitem[{Reid {et~al.}(2009)}]{reid09}
Reid, M.~J., {et~al.} 2009, \apj, 700, 137

\bibitem[{Rocha-Pinto} {et al.}(2004)]{rocha04}
Rocha-Pinto, H. J.,Majewski, S. R., Skrutskie, M. F., Crane, J. D., \& Patterson, R. J. 2004, \apj, 615, 732

\bibitem[{{Robin} {et~al.}(2003){Robin}, {Reyl{\'e}}, {Derri{\`e}re}, \&
  {Picaud}}]{besancon}
{Robin}, A.~C., {Reyl{\'e}}, C., {Derri{\`e}re}, S., \& {Picaud}, S. 2003,
  \aap, 409, 523

\bibitem[{{Sales} {et~al.}(2007){Sales}, {Navarro}, {Abadi}, \&
  {Steinmetz}}]{sales07}
{Sales}, L.~V., {Navarro}, J.~F., {Abadi}, M.~G., \& {Steinmetz}, M. 2007,
  \mnras, 379, 1464

\bibitem[{Samurovi{\'c}} \& {Lalovi{\'c}}(2011)]{samurovic11}
{Samurovi{\'c}}, S., \& {Lalovi{\'c}}, A. 2011, \aap, 531, 82

\bibitem[{{Sch{\"o}nrich}(2012)}]{schonrich12}
{Sch{\"o}nrich}, R. 2012, \mnras, 427, 274

\bibitem[{{Sch{\"o}nrich} {et~al.}(2011){Sch{\"o}nrich}, {Asplund}, \&
  {Casagrande}}]{schonrich11}
{Sch{\"o}nrich}, R., {Asplund}, M., \& {Casagrande}, L. 2011, \mnras, 415, 3807

\bibitem[{{Sch{\"o}nrich} {et~al.}(2010){Sch{\"o}nrich}, {Binney}, \&
  {Dehnen}}]{schonrich10}
{Sch{\"o}nrich}, R., {Binney}, J., \& {Dehnen}, W. 2010, \mnras, 403, 1829

\bibitem[{{Sesar} {et~al.}(2011){Sesar}, {Juri{\'c}}, \&
  {Ivezi{\'c}}}]{sesar11}
{Sesar}, B., {Juri{\'c}}, M., \& {Ivezi{\'c}}, {\v Z}. 2011, \apj, 731, 4

\bibitem[{Sirianni {et~al.}(2005)}]{sirianni05}
Sirianni, M., {et~al.} 2005, \pasp, 117, 1049

\bibitem[{Sirko {et~al.}(2004)}]{sirko04}
Sirko, E., {et~al.} 2004, \aj, 127, 914

\bibitem[{Smith {et~al.}(2009)}]{smith09}
Smith, M.~C., {et~al.} 2009, \mnras, 399, 1223

\bibitem[{{Sohn} {et~al.}(2012a){Sohn}, {Anderson}, \& {van der
  Marel}}]{sohn12a}
{Sohn}, S.~T., {Anderson}, J., \& {van der Marel}, R.~P. 2012a, \apj, 753, 7

\bibitem[{{Sohn} {et~al.}(2012b){Sohn}, {Besla}, {van der Marel},
  {Boylan-Kolchin}, {Majewski}, \& {Bullock}}]{sohn12b}
{Sohn}, S.~T., {Besla}, G., {van der Marel}, R.~P., {Boylan-Kolchin}, M.,
  {Majewski}, S.~R., \& {Bullock}, J.~S. 2012b, ApJ submitted,
  (arXiv:1210.6039)

\bibitem[{{van der Marel} {et~al.}(2012b){van der Marel}, {Besla}, {Cox},
  {Sohn}, \& {Anderson}}]{vandermarel12b}
{van der Marel}, R.~P., {Besla}, G., {Cox}, T.~J., {Sohn}, S.~T., \&
  {Anderson}, J. 2012b, \apj, 753, 9

\bibitem[{{van der Marel} {et~al.}(2012a){van der Marel}, {Fardal}, {Besla},
  {Beaton}, {Sohn}, {Anderson}, {Brown}, \& {Guhathakurta}}]{vandermarel12a}
{van der Marel}, R.~P., {Fardal}, M., {Besla}, G., {Beaton}, R.~L., {Sohn},
  S.~T., {Anderson}, J., {Brown}, T., \& {Guhathakurta}, P. 2012a, \apj, 753, 8

\bibitem[{{VandenBerg} {et~al.}(2006){VandenBerg}, {Bergbusch}, \&
  {Dowler}}]{vandenberg06}
{VandenBerg}, D.~A., {Bergbusch}, P.~A., \& {Dowler}, P.~D. 2006, \apjs, 162,
  375

\bibitem[{Watkins {et~al.}(2009)}]{watkins09}
Watkins, L.~L., {et~al.} 2009, \mnras, 398, 1757

\bibitem[{Xue {et~al.}(2008)}]{xue08}
Xue, X.~X., {et~al.} 2008, \apj, 684, 1143

\bibitem[{Xue {et~al.}(2011)}]{xue11}
Xue, X.-X., {et~al.} 2011, \apj, 738, 79

\bibitem[{{Zinn}(1985)}]{zinn85}
{Zinn}, R. 1985, \apj, 293, 424

\bibitem[{{Zolotov} {et~al.}(2009){Zolotov}, {Willman}, {Brooks}, {Governato},
  {Brook}, {Hogg}, {Quinn}, \& {Stinson}}]{zolotov09}
{Zolotov}, A., {Willman}, B., {Brooks}, A.~M., {Governato}, F., {Brook}, C.~B.,
  {Hogg}, D.~W., {Quinn}, T., \& {Stinson}, G. 2009, \apj, 702, 1058

\end{thebibliography}

\end{document}